\documentclass[11pt]{article}

\usepackage{amsmath,amsfonts,amssymb,amsthm}
\usepackage{times}
\usepackage[margin=1.2in]{geometry}
\usepackage{graphicx,soul}
\usepackage[margin=6pt,skip=0pt,font=small,labelfont=bf,labelsep=period]{caption}

\newtheorem{theorem}{Theorem}[section]

\newcommand{\heading}[1]{\medskip\noindent{\emph{#1}}}
\newcommand{\lmax}{\Lambda_{\max}}
\newcommand{\R}{\mathbb R}

\newcommand{\eps}{\varepsilon}

\newcounter{kevinslistcountertoo}%
\renewenvironment{enumerate}{
  \begin{list}{(\roman{kevinslistcountertoo})}
    {\usecounter{kevinslistcountertoo}
      \setlength{\parsep}{3pt}
      \setlength{\labelwidth}{24pt}
      \setlength{\itemsep}{1pt}
      \setlength{\topsep}{3pt}}}{\end{list}}

\begin{document}

\title{Limitations of Perturbative Techniques in the Analysis of
  Rhythms and Oscillations}

\author{Kevin K.~Lin\thanks{Department of Mathematics and Program in
    Applied Mathematics, University of Arizona, USA}, Kyle
  C.~A.~Wedgwood\thanks{School of Mathematical Sciences, University of
    Nottingham, UK}, Stephen Coombes$^\dagger$, Lai-Sang
  Young\thanks{Courant Institute of Mathematical Sciences, New York
    University, USA}}

\date{\today}

\maketitle

\begin{abstract}
  Perturbation theory is an important tool in the analysis of
  oscillators and their response to external stimuli.  It is predicated
  on the assumption that the perturbations in question are
  ``sufficiently weak", an assumption that is not always valid when
  perturbative methods are applied. In this paper, we identify a number
  of concrete dynamical scenarios in which a standard perturbative
  technique, based on the infinitesimal \emph{phase response curve}
  (PRC), is shown to give different predictions than the full model.
  Shear-induced chaos, {\it i.e.}, chaotic behavior that results from
  the amplification of small perturbations by underlying shear, is
  missed entirely by the PRC. We show also that the presence of
  ``sticky'' phase-space structures tend to cause perturbative
  techniques to overestimate the frequencies and regularity of the
  oscillations.  The phenomena we describe can all be observed in a
  simple 2D neuron model, which we choose for illustration as the PRC is
  widely used in mathematical neuroscience.
\end{abstract}



\section*{Introduction}

Rhythmic activity is commonplace in biological phenomena: the
spontaneous beating of heart cells in culture \cite{glass84}, the
synchronization of flashing fireflies \cite{mirollo-strogatz}, and
central pattern generators in animal locomotion \cite{cohen-cpg}, and
calcium oscillations that underlie a plethora of cellular responses
(ranging from muscle contraction to neurosecretion) \cite{thul} are just
a few examples (see, e.g., \cite{GM} and \cite{winfree} for many more).
Mathematical models of biological oscillations often provide useful
insights into the underlying biological process; for example, they can
explain the observed robustness of circadian rhythms \cite{winfree} and
of population cycles \cite{may}, and can be used to infer plausible
structures for central pattern generator networks based on locomotion
gaits \cite{golubitsky}.
Analyzing models of biological oscillations, however, is generally not
easy: the mechanisms underlying biological rhythms are varied and
complex, and this complexity is reflected in the corresponding
mathematical models.  Indeed, models of biological oscillators are often
high dimensional, highly nonlinear, and have uncertain parameters, all
of which make them challenging to study.

Perturbation theory, long a staple of applied mathematics, provides a
practical solution in many situations.  Mathematically, robust
oscillations correspond to attracting limit cycles in phase space.  If
the stimuli involved are not too strong, then one is justified in
viewing trajectories of the forced system as perturbations of the
original limit cycle.  When judiciously applied, such perturbative
analyses can yield a great deal of insight and useful quantitative
predictions.  For oscillators, a good example of an effective
perturbation technique is that of infinitesimal phase response curves
(PRCs) and associated phase reductions \cite{ermkop}.  The infinitesimal
PRC of an oscillator records the {\em phase change} that results from an
applied perturbation.
Given an oscillator, there are many ways to obtain its infinitesimal
PRC: in addition to analytical perturbation techniques, there are
efficient numerical methods for constructing PRCs; the entire PRC itself
can even be inferred directly from experimental data.
Moreover, PRC-based techniques require only tracking just the phase of
the oscillator, thereby greatly reducing complexity.
They are used in many areas of mathematical biology, but especially in
computational neuroscience, as they yield direct predictions about the
modulation of spike timing and frequency by external stimuli.
Furthermore, they allow one to make predictions for weakly-coupled
networks \cite{hoppensteadt-izhikevich,piko01}~.

However, as is well known, perturbation methods do not always correctly
reflect the true dynamical picture, as they systematically overlook
certain aspects of the dynamics.  In using perturbation theory, one {\it
  assumes} that the perturbation is small, an assumption that is not
always valid in applications.  In this paper, we identify a number of
concrete scenarios in which PRCs and phase reductions give predictions
different from that of the full model.  One situation is when the
perturbation causes the trajectory to leave the basin of attraction of
the limit cycle, which can occur even with moderately weak forcing.  But
even without leaving the basin, more subtle effects can lead
perturbation theory astray, giving --- to varying degrees --- incorrect
predictions.  These situations include the presence of ``sticky''
invariant phase-space structures near a limit cycle, which can cause
perturbation theory to overestimate the regularity and frequency of a
stimulated oscillator.  We will also show that dynamical shear in a
neighborhood of the oscillator can cause it to behave chaotically when
forced.  These scenarios cannot be captured by infinitesimal phase
reductions.

The scenarios described in this paper are relevant for general
oscillators, but in view of the popularity of PRC-based techniques in
neuroscience, we will illustrate our ideas using the Morris-Lecar (ML)
neuron model~\cite{ER,ET}.  This widely-used model provides a convenient
and flexible example because of its low dimensionality and rich
bifurcation structure.  The phenomena of interest are not hard to find
in certain ML regimes; they do not require stringent tuning of parameters.
We point out also that while we focus
here on single neuron models, our findings remain relevant for
oscillators operating within networks.

\medskip

This paper is organized as follows: In Sect.~\ref{sect:bkgnd}, we review
some relevant mathematical background, including brief discussions of phase
response theory and ``shear-induced chaos", a general
mechanism for producing chaotic behavior in driven oscillators.
Sect.~\ref{sect:neurosci} introduces the ML model and some relevant
ideas from computational neuroscience.  
In the last three sections, 
certain regimes of the ML model are used to demonstrate how perturbative
techniques sometimes do not correctly predict the behavior of the full model. 
Sect.~\ref{sect:shear} contains an example in which the infinitesimal PRC
gives no hint of the strange attractor in the full model.
Sects.~\ref{sect:stuck} and \ref{sect:bistable} 
illustrate how the presence of nearby invariant structures can impact
neuronal response in ways that cannot be captured by PRCs alone.

\section{Mathematical background}
\label{sect:bkgnd}

The general setting for this section is a nonlinear oscillator modeled
by an $n$-dimensional ODE $\dot{x} = f(x)$ with a limit cycle $\gamma$~.
We assume throughout that $\gamma$ is not only attractive as a periodic
orbit but {\em hyperbolic}, i.e., its Floquet multipliers have absolute
values $<1$.  The period of $\gamma$ is denoted ${\rm Per}(\gamma)$.

\subsection{Phase response curves and phase reductions}
\label{sect:prc}

This section contains a brief review of a perturbation technique for 
oscillators known as (infinitesimal) phase response theory.  
For more details and many applications, see, e.g.,~\cite{ET,GM,guck75,winfree}.

First, we fix a notion of ``phase" on the oscillator: Fix a reference
point $x_*\in\gamma$, and declare its phase to be
$\psi(x_*)=0$.\footnote{In neuroscience, it is customary to define zero
  phase to be the moment the neuron ``spikes.''}  For any other point
$x\in\gamma$, the phase $\psi(x)$ can be defined to be the amount of
time to go from $x_*$ to $x$ along the cycle $\gamma$~.  Next we extend
this notion to a neighborhood of $\gamma$. It is a mathematical fact
that $\psi$ extends uniquely to a smooth function such that for all
solutions $x$ with $x(0)$ near $\gamma$, $\frac{d}{dt}\psi(x(t)) = 1$~.
Thus $\psi$ serves as a kind of ``clock'' for tracking the passage of
time along trajectories near $\gamma$.

Our goal is to describe what happens to the phase when a time-dependent
perturbation is applied.  Let $\theta(t) := \psi(x(t))$, where $x$ is a
perturbed trajectory.  We would like an approximate equation for
$\theta(t)$~.  Rather than doing this for general perturbations, we
consider a perturbed equation of the form
\begin{equation}
  \dot{x} = f(x) + I(t)\hat{k}~,
  \label{eq:ode+forcing}
\end{equation}
where $x(t)\in\R^n$~, $I$ is a scalar signal, and $\hat{k}\in\R^n$ is a
constant vector.  (In neuroscience applications, for example, one of the
variables usually represents the membrane voltage, and this is typically
the only variable that can be directly affected by external
perturbations.) Now let $\xi:\R\to\R^n$ be a periodic solution to the
``adjoint equation''
\begin{equation}
  \dot{\xi}(t) = -Df(\gamma(t))^T\cdot \xi(t)~,
  \label{eq:adjoint}
\end{equation}
where $\gamma(t)$ denotes an orbit parametrizing the cycle $\gamma$~,
and let $\Delta(\theta) := \xi(\theta)\cdot\hat{k}$~.  Under the
normalization condition $\xi(0)\cdot f(0) =1$,\footnote{It is easy to check that if this
  condition holds for $t=0$, it holds for all $t$.} 
   it can be shown (see, e.g., \cite{ET}) that if $x$ is a solution of
Eq.~(\ref{eq:ode+forcing}) and $I=O(\eps)$ for a small parameter $\eps$,
then $\theta$ satisfies
\begin{equation}
  \dot{\theta} = 1 + \Delta(\theta) I(t) + O(\eps^2)~.
  \label{eq:phase-reduction}
\end{equation}
Truncating all terms of $O(\eps^2)$ in Eq.~(\ref{eq:phase-reduction})
yields an equation for $\theta$, the {\em phase reduction} of
Eq.~(\ref{eq:ode+forcing}).  The function $\Delta$ is the {\em
  infinitesimal phase response curve} (PRC).\footnote{Some authors refer
  to $\Delta$ as the phase resetting curve.}

For systems that are near a bifurcation, the above procedure can be used
to derive analytical approximations of the PRC via normal forms
\cite{brown-moehlis}.  In more general situations (where there are few
practical analytical techniques), one can obtain PRCs through numerical
computation, or even directly from experimental measurements (see, e.g.,
\cite{oprisan,netoff,winfree} and references therein).  This flexibility
and accessibility is part of its appeal in mathematical biology, and
in neuroscience in particular.

\heading{Stochastic forcing.}  The basic methodology of infinitesimal
phase response curves can be extended to systems driven by stochastic
forcing.  That is, suppose that in addition to a deterministic forcing,
we add a second white-noise term:
\begin{displaymath}
  \dot{x} = f(x) + (I(t) +
  \beta\dot{W})~\hat{k}~;\qquad\dot{W}=\mbox{white noise.}
\end{displaymath}
In \cite{LE}, Ly and Ermentrout show that with the above forcing,
$\theta$ satisfies the equation
\begin{equation}
  \dot{\theta} = 1 + \Delta(\theta) I(t) +
  \frac{\beta^2}{2}\Delta(\theta)\Delta'(\theta) +
  \beta\Delta(\theta)\dot{W}~~+~~(\mbox{higher order terms})~,
  \label{eq:prc-sde}
\end{equation}
assuming both $\eps$ and $\beta$ are small.

Eq.~(\ref{eq:prc-sde}) can be used to derive a number of quantities of
interest.  For example, if we view
Eq.~(\ref{eq:prc-sde}) as modeling a neuron that ``fires'' whenever
$\theta=0$~, then a result of Ly and Ermentrout states that the firing
rate resulting from a constant forcing $I(t)\equiv\eps$ is
\begin{equation}
  r(\eps) = 1 + \eps\bar\Delta +
  \eps^2\int_0^{{\rm Per}(\gamma)}\Big({\bar\Delta}^2 -
  \Delta^2(\theta)\Big)~d\theta +
  \frac{\beta^4}{4}\int_0^{{\rm Per}(\gamma)}\Delta^2(\theta)\big(\Delta'(\theta)\big)^2~d\theta
  + O(\eps^3)~,
  \label{eq:prc-firing-rate}
\end{equation}
to leading order in $\beta$~.  In the above, $\bar\Delta =
\int_0^{{\rm Per}(\gamma)}\Delta(\theta)~d\theta$.

\bigskip
\heading{Remark on terminology.}  Other variants of the PRC exist for
finite-size perturbations.  In this paper, the term ``PRC'' will always
mean the {\em infinitesimal} phase response curve.

\subsection{When do phase reductions work, and when might they fail?}
\label{sect:when-prc-works}

Because the PRC is defined in terms of a vector field $f$ and its
derivative along a limit cycle $\gamma$ (see Eq.~(\ref{eq:adjoint})), it
can only contain information about the flow in an infinitesimal
neighborhood of $\gamma$.  We discuss briefly in this subsection a few
scenarios in which the behavior of the flow a finite distance away from
$\gamma$ can have a dramatic effect on the oscillator's response. These
ideas are illustrated in concrete examples in Sects.~\ref{sect:shear} --
\ref{sect:bistable}.

\medskip
\noindent
(1) {\em Leaving the basin of $\gamma$}.  The simplest possible way for
something to go wrong is when the forcing carries a phase point outside
of the basin of attraction of $\gamma$.  If $\Phi_t$ denotes the
unforced flow, then the basin of $\gamma$, denoted ${\rm
  Basin}(\gamma)$, is defined to be the set of all phase points $x$ such
that $\Phi_t(x) \to \gamma$ as $t \to \infty$. Once a perturbation, say
in the form of a ``kick", causes a trajectory to leave ${\rm
  Basin}(\gamma)$, what it does may depend on dynamical structures far
away from $\gamma$.  For example, if the system is bistable, or
multi-stable, i.e., it has more than one attracting set, which can be in
the form of a stationary point, a limit cycle, or something more
complicated, then an ``escaped" trajectory can end up near one of these
structures, and in time, it may -- or may not -- get kicked back into
${\rm Basin}(\gamma)$.  Needless to say, the behavior of such a
trajectory bears little resemblance to that predicted by the PRC. This
scenario must be taken into consideration when the forcing is strong
relative to the distance of $\partial({\rm Basin}(\gamma))$ to $\gamma$.
(Here $\partial({\rm Basin}(\gamma))$ is the boundary of ${\rm
  Basin}(\gamma)$.)

\medskip
\noindent
(2) {\em Invariant structures and ``trapping"}.  Even without venturing
outside of ${\rm Basin}(\gamma)$, a perturbed trajectory that comes near
$\partial({\rm Basin}(\gamma))$ can be nontrivially affected by certain
dynamical structures in $\partial({\rm Basin}(\gamma))$.  These
structures may seem innocuous -- they are non-attracting -- but as we
will see, they can seriously impact the surrounding dynamics. Consider,
for example, a saddle fixed point of the unperturbed flow. An orbit that
comes near it will, under the unperturbed flow, remain near it for some
duration of time (depending on the ratios of the eigenvalues of the
linearized flow at that point). Since an invariant structure is
typically not truly invariant for the perturbed flow, a perturbed
trajectory that gets near it will likely escape eventually and return to
$\gamma$.  However, the escape time can be quite long, and this effect
cannot be captured by phase reductions. The tendency to remain near
invariant structures can be mitigated by the forcing when the forcing
acts to push trajectories away; by the same token, it can also be
magnified if the forcing ``conspires" to keep trajectories in a
region. Indeed, there is no reason why a forcing cannot -- by itself --
create trapping regions within the basin of $\gamma$ if the attraction
to $\gamma$ is weak compared to the forcing.

\begin{figure}
  \begin{center}
    \includegraphics[bb=0 0 307 117,scale=0.75]{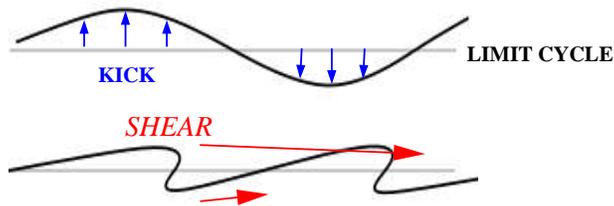}
  \end{center}
  \caption{The stretch-and-fold action of a kick followed by
    relaxation in the presence of shear.}
  \label{shear}
\end{figure}

\medskip
\noindent
(3) {\em Shear-induced chaos}.  This is yet another dynamical phenomenon
that cannot be captured by (infinitesimal) PRCs. This phenomenon is
illustrated in Fig.~\ref{shear}.  In each of the two pictures, $\gamma$
is represented by the horizontal line. Shear refers to the differential
in the horizontal component of the velocity as one moves vertically up
in the phase space; here points above $\gamma$ move around $\gamma$
faster than points below.  Suppose that an impulsive perturbation, or
``kick,'' applied in such a situation produces a ``bump" in $\gamma$. As
the flow relaxes, this ``bump" is attracted back to the original limit
cycle. As it evolves, it is folded and stretched by the flow if
sufficient shear is present, as one can visualize in Fig.~\ref{shear}.
Since stretching and folding of phase space is associated with complex
dynamical behavior such as horseshoes and strange attractors \cite{gh}, this
picture suggests that perturbing a limit cycle with strong shear can
lead to chaotic dynamics.  That this can indeed happen has been
established (rigorously) in recent developments in dynamical systems
theory.

\medskip
To connect paragraphs (2) and (3), we mention that invariant structures
in $\partial({\rm Basin}(\gamma))$ can be a contributing factor to
shear, but shear can also arise for many other reasons.  Since the
results on shear-induced chaos alluded to above are not widely known in
the mathematical biology community, we will provide a more detailed
review in the next subsection.

\subsection{Shear-induced chaos and related phenomena}
\label{wang-young}

In this section, we review some of the geometric ideas put forth in
\cite{WY2,WY3} (and also \cite{WY1,WY4} and \cite{OS}).  The exposition
here roughly follows \cite{LY-smale}, which contains a more thorough
(but still non-technical) discussion.  We focus here on {\em
  periodically-kicked oscillators,} since in this setting the various
dynamical mechanisms are most transparent.

A periodically-kicked oscillator
is a system of the form
\begin{equation}
  \dot{x} = f(x) + A\cdot H(x)\sum_{k=-\infty}^\infty \delta(t-kT)~,
  \label{eq:periodic-kicks}
\end{equation}
where $(A,T)\in\R\times\R^+$ are parameters and $H:\R^n\to\R^n$ is a
given smooth function.  We assume as before that $\dot{x}=f(x)$ has an
attracting hyperbolic limit cycle $\gamma$~.
Eq.~(\ref{eq:periodic-kicks}) thus models an oscillator that is given a
sharp ``kick'' every $T$ units of time.  We interpret the kicks as
follows: whenever $t=nT$, we apply a mapping $\kappa$ (defined by $H$)
to the system; between kicks, the system follows the flow $\Phi_t$
generated by $\dot{x}=f(x)$~.  When kicks are applied repeatedly, the
dynamics of Eq.~(\ref{eq:periodic-kicks}) can be captured by iterating
the time-$T$ map $F_T = \Phi_T \circ \kappa$.  If there is a
neighborhood $\cal U$ of $\gamma$ such that $\kappa({\cal U}) \subset
{\rm Basin}(\gamma)$, and $T$ is long enough that points in
$\kappa({\cal U})$ return to $\cal U$, {\it i.e.}, $F_T({\cal U})
\subset {\cal U}$, then $\Gamma = \cap_{n \geq 0} F_T^n({\cal U})$ is an
attractor for the periodically kicked system $F_T$.  One can view
$\Gamma=\Gamma(\kappa, T)$ as what becomes of the limit cycle $\gamma$
when the oscillator is periodically kicked.

The structure of $\Gamma$ and the associated dynamics depends strongly
on the kick parameters $A$ and $T$, as well as on the relation between
the kick map $\kappa$ and the flow near $\gamma$~.  When $A$ is small,
we generally expect $\Gamma$ to be a slightly perturbed version of
$\gamma$~, because (as is well known) hyperbolic limit cycles are robust
under small perturbations~\cite{gh}.  In this case, $\Gamma$ is known as
an {\em invariant circle,} and the restriction of $F_T$ to $\Gamma$ is
equivalent to a diffeomorphism on $S^1$~.  Circle diffeomorphisms are
well known to exhibit essentially two distinct types of behavior: {\em
  quasiperiodic motion,} in which the mapping is equivalent to rotation
by an irrational angle, and {\em gradient-like} behavior characterized
by sinks and sources on the invariant circle.  In terms of the kicked
oscillator dynamics, the former corresponds to the driven oscillator
drifting in and out of phase relative to the kicks, while the latter
corresponds to stable phase-locking.

The preceding discussion suggests that when kicks are weak, we should
expect fairly regular behavior.  To obtain more complicated behavior, it
is necessary to ``break'' the invariant circle.  The main idea is best
illustrated in the following {\em linear shear model}, a version of
which was first studied in \cite{zaslav}:

\begin{equation}
  \begin{array}{rcl}
    \dot{\theta} & =& 1 + \sigma y\\[1ex]
    \dot{y} &=& -\lambda y + A\cdot H(\theta) \cdot\sum_{n=0}^\infty \delta(t-nT)
  \end{array}
  \label{eq:linear-shear}
\end{equation}
where $(\theta, y) \in S^1 \times {\mathbb R}$ are coordinates in the
phase space, $\lambda$, $\sigma$, and $A$ are constants, and $H: S^1 \to
{\mathbb R}$ is a {\rm non-constant} smooth function.  When $A=0$, the
unforced system has a limit cycle $\gamma = S^1\times\{0\}$.  The
following result, due to Wang and Young, shows that
Eq.~(\ref{eq:linear-shear}) indeed exhibits chaotic behavior under the
right conditions.  (There is an obvious analog in $n$-dimensions
\cite{WY3}.)

\begin{theorem}  {\rm\cite{WY3}}
  \label{thm:linear-model}
  Consider the system in Eq.~(\ref{eq:linear-shear}).  If the quantity
  $$
  \frac{\sigma}{\lambda} \cdot A \ \equiv \
  \frac{\mbox{\rm shear}}{\mbox{\rm contraction rate}} \cdot
  (\mbox{\rm kick ``amplitude''})
  $$ is sufficiently large (how large depends on the forcing function
  $H$), then there is a positive measure set $\Delta \subset {\mathbb
    R}^+$ such that for all $T \in \Delta$, $\Gamma$ is a ``strange
  attractor'' of $F_T$~.
\end{theorem}
It is important that $H$ be non-constant, as $H(\theta)$ is what creates
the bumps in Fig.~\ref{shear}.  The geometric meaning of the term
involving $\sigma$, the {\it shear}, is as depicted in Fig.~\ref{shear}.
It is easy to see why $ \frac{\sigma}{\lambda}\cdot A$ is key to
production of chaos by fixing two of these quantities and varying the
third: the larger $\sigma$, the larger the fold; the same is true for
larger kick size $A$.  Notice also that weaker limit cycles are more
prone to shear-induced chaos: the closer $\lambda$ is to $0$, the slower
$\kappa(\gamma)$ returns to $\gamma$, and the longer the shear acts on
it, assuming $T$ is large enough.

The term ``strange attractor'' in Theorem~\ref{thm:linear-model} is used
as short-hand for an attractor with an {\em SRB measure},\footnote{For
  more information, see \cite{eckrue,lsy-srb}.} which roughly speaking,
implies that the trajectory is unstable, or has a positive Lyapunov
exponent, starting from Lebesgue-almost every initial condition in the
basin of the attractor (or at least in a positive measure set).  We say
such a system has a ``strange attractor" because it exhibits {\em
  sustained, observable chaos}, i.e., chaotic behavior that is sustained
in time, and observable for large sets of initial conditions. This is a
considerably stronger form of chaos than the presence of horseshoes
alone (see e.g. \cite{gh} for a discussion of horseshoes).  In the
latter, it is possible for almost all orbits to head toward stable
equilibria, resulting in negative Lyapunov exponents; this scenario, in
which horseshoes coexist with sinks, is known as {\em transient chaos.}%
\footnote{Note that Theorem~\ref{thm:linear-model} asserts the existence
  of ``strange attractors'' only for a {\it positive measure set} of
  $T$, not for all large $T$~.  Indeed, there exist arbitrarily large
  $T$ in the complement of $\Delta$ for which $F_T$ exhibits only
  transient chaos.}

\medskip
\heading{Shear-induced chaos and the geometry of strong stable manifolds
or isochrons}  

\medskip
\noindent
We now return to the general setting of Eq. (\ref{eq:periodic-kicks})
and seek to understand what plays the role of the shear (as $\sigma$ is
no longer defined). Let $\gamma$ and $\Phi_t$ be as at the beginning of
Sect.~\ref{wang-young}. Crucial to this understanding is the following
dynamical structure of the unperturbed flow $\Phi_t$~: For $x \in
\gamma$, define the {\it strong stable manifold} or {\it isochron}
through $x$ to be
$$
W^{ss}(x) = \{y : |\Phi_t(y) - \Phi_t(x)|\to0 \ {\rm as} \  t\to\infty \}\ .
$$ With $\gamma$ assumed to be hyperbolic, it is known that (see, e.g.,
\cite{guck75})
\begin{enumerate}

\item $W^{ss}(x)$ intersects $\gamma$ transversally at exactly one
  point, namely $x$~, and these manifolds are invariant in the sense
  that $\Phi_t(W^{ss}(x)) =W^{ss}(\Phi_t(x))$~.
  
\item $\{W^{ss}(x), x \in \gamma\}$ partitions ${\rm Basin}(\gamma)$
  into codimension-1 submanifolds.

\end{enumerate}

\begin{figure}
  \begin{center}
    \includegraphics[bb=0 0 379 107,scale=0.75]{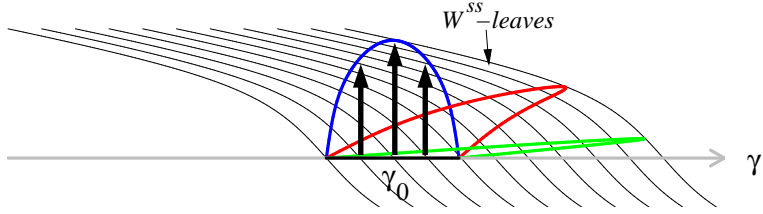}
  \end{center}
  \caption{Geometry of folding in relation to the $W^{ss}$-foliation.
    Shown are the kicked image of a segment $\gamma_0$ and two of its
    subsequent images under $\Phi_{np}$~, $p={\rm Per}(\gamma)$.}
  \label{wss}
\end{figure}

We now examine the action of the kick map $\kappa$ in relation to the
$W^{ss}$-foliation.  Fig.~\ref{wss} is analogous to Fig.~\ref{shear}; it
shows the image of a segment $\gamma_0$ of $\gamma$ under $F_T = \Phi_T
\circ \kappa$. For illustration purposes, we assume $\gamma_0$ is kicked
upward with its end points held fixed, and assume $T=np$ for some $n \in
{\mathbb Z}^+$ (otherwise the picture is shifted to another part of
$\gamma$ but is qualitatively similar). Since $\Phi_{np}$ leaves each
$W^{ss}$-manifold invariant, we may imagine that during relaxation, the
flow ``slides'' each point of the curve $\kappa(\gamma_0)$ back toward
$\gamma$ along $W^{ss}$-manifolds.  In the situation depicted, the
effect of the folding is evident.

Fig.~\ref{wss} gives considerable insight into what types of kicks are
conducive to the formation of strange attractors.  Kicks along
$W^{ss}$-manifolds or in directions roughly parallel to the
$W^{ss}$-manifolds will not produce strange attractors, nor will kicks
that essentially carry one $W^{ss}$-manifold to another.  What causes
the stretching and folding is the {\it variation} in how far points
$x\in\gamma$ are moved by $\kappa$ as measured in the direction
transverse to the $W^{ss}$-manifolds.  In the simple model of
Eq.~(\ref{eq:linear-shear}), because of the linearity of the unforced
equation, $W^{ss}$-manifolds are straight lines with slope
$~-\lambda/\sigma$ in $(\theta, y)$-coordinates.  Variation in the sense
above is created by any non-constant $H$; the larger the ratio
$\frac{\sigma}{\lambda}A$, the greater this variation.

\medskip

The notion of ``phases" in Sect.~\ref{sect:prc} is defined precisely by
the partition of neighborhoods of $\gamma$ into $W^{ss}$-manifolds or
isochrons (two names used by different communities for the same object),
i.e., we view $x \in {\rm Basin}(\gamma)$ as having the same phase as $y
\in \gamma$ if $x \in W^{ss}(y)$.  The ideas in the last paragraph are
in the same spirit as the {\em phase transition curves} introduced by
Winfree \cite{winfree,GM,ET}.  
They were discovered independently in the
rigorous work of Wang and Young, who proved, under suitable geometric
conditions on phase variations, the existence of strange attractors
having many of the properties commonly associated with chaos 
\cite{WY1,WY4}. These ideas have since been applied to various situations;
rigorous results include \cite{WY2,WY3,LWY,OS,WO} and \cite{lee-sri},
and numerical results indicate the occurrence of shear-induced chaos 
in broader dynamical settings \cite{LY-nonlin,LSY12,Lin-HH}.

\subsection{Summary and comparisons}
\label{sect:going-forward}

Sects.~\ref{sect:prc} and \ref{wang-young} outlined two seemingly
distinct approaches to the analysis of phase response.  These approaches
are in fact closely related: In Sect.~\ref{sect:prc}, perturbations are
assumed to be small, which geometrically means one can approximate
$W^{ss}(y)$ by its linearization around $\gamma$ . The procedure of
phase reduction then amounts to viewing the perturbation as a sequence
of infinitesimal kicks, projecting the kicked trajectory back to
$\gamma$ along $W^{ss}$-leaves following each kick.  The analysis
sketched in Sect.~\ref{wang-young} is a more global version of the same
idea: here the perturbed orbit is allowed to wander farther away from
$\gamma$, and one takes into consideration the geometry (or curvature)
of the $W^{ss}$-manifolds in relation to the kick in assessing its
impact.

A notable difference between full-model analyses and phase reductions is
that the latter rule out {\it a priori} any possibility of chaotic
behavior.  As explained in Sect.~\ref{wang-young}, in the full model,
fairly innocuous kicks applied ``the right way" can lead to positive
Lyapunov exponents for large sets of initial conditions.  This cannot be
captured by the infinitesimal PRC, for flows in one spatial dimension
are never chaotic.

With regard to practicalities, full model analyses are, needless to say, more costly. 
While our aim here has been
to raise awareness of the issues discussed in
Sects.~\ref{sect:when-prc-works} and \ref{wang-young}, it
is not our intention to advocate that one necessarily
starts by computing $W^{ss}$-manifolds (even though they are computable
in many situations).  In many cases, by far the most direct way to get a
quick idea of whether shear-induced chaos is present is to look at
$F_T$-images of $\gamma$ (recall that $F_T = \Phi_T \circ \kappa$ is the
composite map obtained by first kicking then following the unperturbed
dynamics for time $T$) and see if folds develop.  See Section 3.

\section{A neuron model}
\label{sect:neurosci}

In Sects.~\ref{sect:shear}--\ref{sect:bistable}, a few concrete scenarios will be presented to illustrate
how and why the infinitesimal PRC may give incorrect predictions. 
The model we use is taken from neuroscience; it is the Morris-Lecar (ML) model
of neuron dynamics. A brief introduction of the ML model and some relevant
information is given below to provide context for readers not familiar with the subject. 
We remark that perturbation methods 
and the infinitesimal PRC in particular are widely used in neuroscience, 
both in the study of single neuron dynamics (see e.g. \cite{ET})
and in the analysis of neuronal networks modeled as
systems of coupled phase oscillators (see e.g. \cite{hoppensteadt-izhikevich}).

\nopagebreak
\heading{The Morris-Lecar (ML) model}  

\medskip
\noindent
The dominant mode of
communication between neurons is via the generation and transmission of
action potentials, or ``spikes'' \cite{dayan-abbott}.  Neurons
accomplish this through the coordinated activity of voltage-sensitive
ion channels in the cell membrane, which open and close in specific ways 
in response to changes in membrane voltage.
The ML model is a simple model of this spike-generation
process for a single neuron.  
It has the form
\begin{equation}
  \begin{array}{rcl}
    C_m~\dot{v} &=& I(t) - g_{\rm leak}\cdot(v-v_{\rm leak}) -
    g_K~w\cdot(v-v_K) - g_{\rm Ca}~m_\infty(v)\cdot(v-v_{\rm Ca}) \\[2ex]
    \dot{w} &=& \phi\cdot\big(w_\infty(v) - w\big)/\tau_w(v)~.
  \end{array}
  \label{eq:ml}
\end{equation}
The variable $v$ is the membrane voltage; the first equation expresses
Kirchoff's current law across the cell membrane, with $I(t)$
representing a stimulus in the form of an incoming current.  The
variable $w$ is a {\em gating variable:} it describes the fraction of
membrane ion channels that are open at any time.  Real-life neurons
typically have multiple kinds of ion channels; in more realistic models
like Hodgkin-Huxley, these are tracked by separate gating variables.
The ML model is simplified in that
there is just
one effective gating variable. 
The forms and meanings
 of the auxiliary functions $w_\infty$~, $\tau_w$~,
$m_\infty$ and other parameters in (\ref{eq:ml}) are given in the Appendix.
For more information we refer the reader to \cite{ET}.

The ML model has a very rich bifurcation structure.  Roughly speaking,
by varying a constant current $I(t)\equiv I_0$~, one observes, in
different parameter regions, dynamical regimes corresponding to sinks,
limit cycles, and Hopf, saddle-node and homoclinic bifurcations, as well
as combinations of the above.  These scenarios together with their
neuroscience interpretations are discussed in detail in \cite{ET}.
We will use two of these scenarios for illustration; relevant features
of these regimes are reviewed as needed in the sections to follow.

\nopagebreak
\heading{Oscillators in neuronal dynamics}

\medskip
\noindent
When a sufficiently large DC current is injected into a neuron, either
artificially or through the action of neurotransmitters, a typical
neuron will begin to spike regularly.  In such situations, one can view
the neuron as an oscillator.  For example, in Eq.~(\ref{eq:ml}), if we
apply a constant driving current $I(t)\equiv I_0$  and slowly increase $I_0$ from 0 to a
large value, the ML neuron will switch from quiescent to spiking at
regular intervals, the latter corresponding to the emergence of a limit cycle.
More generally, if a neuron is operating in a ``mean-driven'' regime, in
which the stimulus it receives consists of a large DC component plus a
(weaker) fluctuating AC component, one can view the AC component of the
stimulus as a perturbation of the oscillator \cite{ET}.

\nopagebreak
\heading{Relevant properties}

\medskip
\noindent
The crudest statistic associated with a spiking neuron it is {\it firing
  rate}, which translates into the {\it frequency} of the neural
oscillator.
Sometimes one is also interested in more refined properties
of spike trains and even the precise timing of spikes.

A neuron or a network of neurons receiving a stimulus is said to be {\em
  reliable} if its response does not vary significantly upon repeated
presentations of the same stimulus.  Reliability is of interest  because it
constrains a neuron's (or network's) ability to encode information via temporal
patterns of spikes.  Mathematically, a stimulus-driven system can be viewed
as a non-autonomous dynamical system of the form $\dot{x} = f(x,
I(t))~,$ where $I(t)$ represents the stimulus. The question of spike-time
reliability, then, boils down to the following: Given a specific signal $\big(I(t) :
t\in[0,\infty)\big)$~, does the response $x(t)$ depend (modulo transients)
in an essential way on $x(0)$, the condition of the system at the onset of
the stimulus?  If the answer is
negative, the system is reliable.  Otherwise, it is unreliable.

The relevant dynamical quantity here is $\lmax$, the largest Lyapunov exponent
of the system~: If $\lmax<0$, then the system is reliable, whereas
$\lmax>0$ leads to unreliability.  Heuristically, this is because
$\lmax<0$ leads to phase space contraction, so that the effects of
initial conditions are quickly forgotten, whereas $\lmax>0$ leads the
system to amplify small differences in initial states.  The reasoning
can be made more precise via the theory of random dynamical systems and
random attractors; see \cite{LSY12,LSY3a,LSY3} for details.

\section{Chaotic response to periodic kicking}
\label{sect:shear}

In this section, we show numerically that shear-induced chaos occurs in
the ``homoclinic regime" of the ML model (see below), leading to a lack of reliability. 
As noted in Sect.~\ref{sect:going-forward}, such a possibility is ruled out {\it a priori} by the 
infinitesimal PRC.

\subsection{Geometry of the ``homoclinic regime"}

We view Eq.~(\ref{eq:ml}) with $I(t) \equiv I_0$ for some fixed $I_0$ 
as the unperturbed system, and apply to it a forcing in the $v$-variable 
(forcing the system in $w$ has no physical meaning).

\begin{figure}
  \begin{center}
    \begin{scriptsize}
      \begin{tabular}{cp{2ex}c}
        \includegraphics[bb=0in 0in 3.5in 3in,scale=0.7]{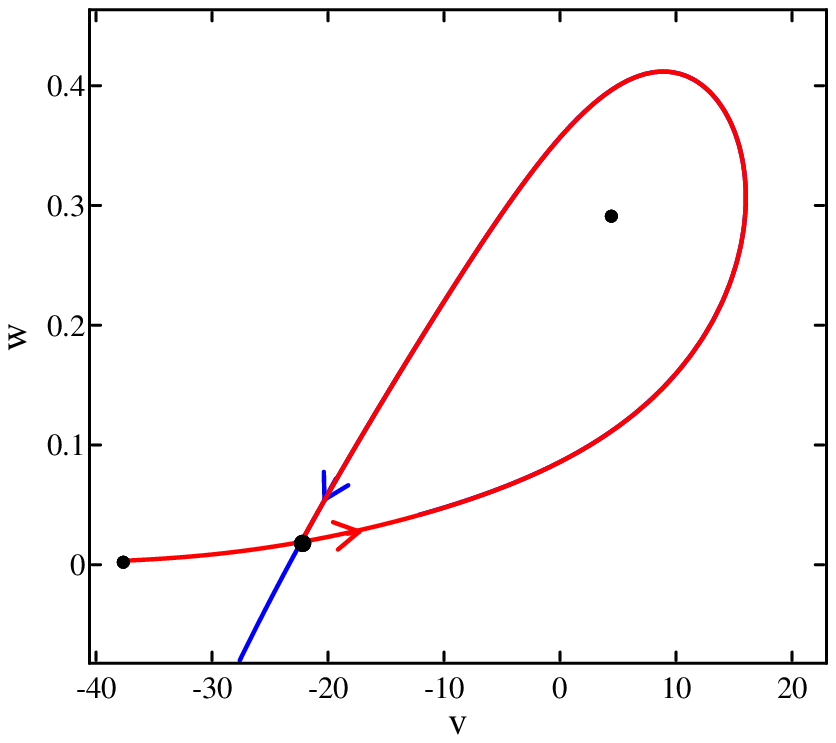} &&
        \includegraphics[bb=0in 0in 3.5in 3in,scale=0.7]{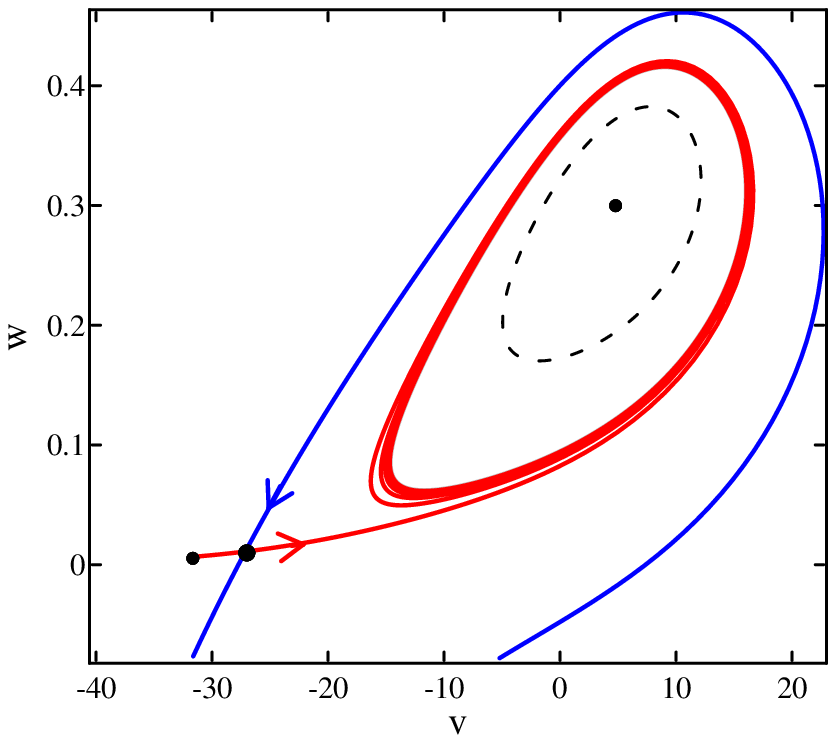}\\
        (a) At the moment of a homoclinic bifurcation, $I_0\approx 35$ &&
        (b) The ``homoclinic regime'' at $I_0=39.5$ \\
      \end{tabular}
    \end{scriptsize}
  \end{center}
  
  \caption{A homoclinic bifurcation and the ``homoclinic regime.''  In
    this parameter regime, the system has 3 fixed points (shown as black
    dots), the middle one being a saddle for a broad range of $I_0$~.
    Panel (a) shows a homoclinic loop anchored at the saddle, with a
    source on the right and a sink on the left.  As $I_0$ increases, the
    homoclinic loop breaks apart and a limit cycle emerges.  Panel (b)
    shows the phase portrait at the parameter regime we use, which is
    well past the homoclinic bifurcation.  In addition to the homoclinic
    bifurcation shown in (a), a (subcritical) Hopf bifurcation has
    occurred, leading the fixed point on the right to become a sink
    surrounded by an unstable periodic orbit / cycle (dashed curve).}
  \label{fig:homoclinic}
\end{figure}

At $I_0 = I_{\rm crit} \approx 35$, for suitable choices of parameters 
(details of which
are given in the Appendix), Eq.~(\ref{eq:ml}) has a homoclinic loop, i.e. there
is a saddle fixed point $p$ one branch of whose stable and unstable manifolds 
coincide (see Fig.~\ref{fig:homoclinic}(a)). To the left of $p$ lies a sink, which attracts the left
branch of the unstable manifold; and there is a source inside the loop.

For $I_0> I_{\rm crit}$, the homoclinic loop is broken, with the unstable manifold  ``inside" the stable manifold. 
Fig.~\ref{fig:homoclinic}(b) shows the phase flow at $I_0=39.5$. 
The unstable manifold wraps around a newly emerged limit cycle, which will
be our $\gamma$. 
Notice also the other dynamical structures: the saddle $p$, its stable
and unstable manifolds and the sink to the left, plus a new sink and unstable
cycle (indicated by the dashed loop) that emerged from the original
source via a Hopf bifurcation (the latter occurs around $I_0\approx36.3$).
In the rest of this section, $I_0$ will be taken around $39.5$, and the
unforced flow is qualitatively similar to that in Fig.~\ref{fig:homoclinic}(b).

We consider periodic kicks applied to such a regime, 
i.e., in Eq.~(\ref{eq:ml}) we take as input $I(t)
= I_0 + A\sum_k \delta(t-kT)~,$ where $|A|$ is the kick amplitude and $T$ the
kick period. Geometrically, this kick corresponds to shifting simultaneously
all phase points by $A$ (which can be positive or negative) in the horizontal
direction. 

If $|A|$ is sufficiently large, some very obvious things can happen:
For example, a large kick to the left can easily drive points 
on $\gamma$ to the left of the stable manifold of the saddle; these points
will then head for the sink on the left and possibly never return.
Another possibility is to drive points into the basin of attraction of
the sink inside the unstable cycle. We will show in the next subsection that something more subtle
can happen
even with small to medium kicks which do not drive points outside of the basin
of $\gamma$. 

\subsection{Shear-induced chaos}
\label{sect:sic-in-ml}

We will proceed in three stages: First, we discuss -- on a theoretical
level -- the dynamical mechanism responsible for shear production 
in the regimes of interest. Then we 
perform some relatively quick, exploratory
 numerical simulations to check that this shear is
sufficiently strong, and to
identify suitable parameters. Finally, we confirm the presence of shear-induced
chaos via careful computations of Lyapunov exponents.

\medskip \noindent
{\em What is the mechanism that produces shear in the present setting?} 

\smallskip\noindent
Take $A<0$,
so that the kick $\kappa$ 
moves the limit cycle $\gamma$ to the left of itself. This pushes
about half of $\gamma$ inside the unstable cycle, and half of it outside and to the left.
Now suppose we go through a period of relaxation, i.e., we apply the ML  flow
$\Phi_t$, no kicks. As points
come down the left side of $\gamma$, those that are closer to the stable
manifold of the saddle $p$ are likely to follow it longer; consequently they will 
come closer to $p$. (We assume the kick is small enough that no point
gets kicked to the other side of the stable manifold.) It is easy to see that the
closer an orbit comes to a saddle fixed point, the longer it remains in its
vicinity -- certainly longer than orbits that are, for example, inside
the unstable cycle. The differential in ``time spent near $p$", if strong enough, may lead
to a fold in the $(\Phi_t \circ \kappa)$-image of $\gamma$.

The argument above can be made rigorous.  Similar ideas have been used
in rigorous work \cite{afraimovich,WO}. 
 But the reasoning is {\it qualitative}:
while it shows that some amount of shear is present, it does not tell us whether
it is sufficient to cause chaos. 

\begin{figure}
  \begin{center}
    \begin{scriptsize}
      \begin{tabular}{ccc}
        $t=0$ & $t=5$ & $t=10$ \\
        \includegraphics[bb=0in 0in 3in 2.25in,scale=0.65]{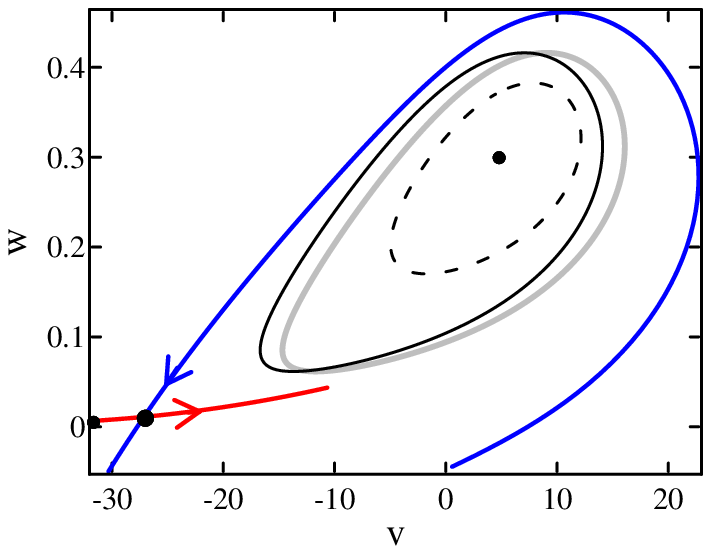} &
        \includegraphics*[bb=0.25in 0in 3in 2.25in,scale=0.65]{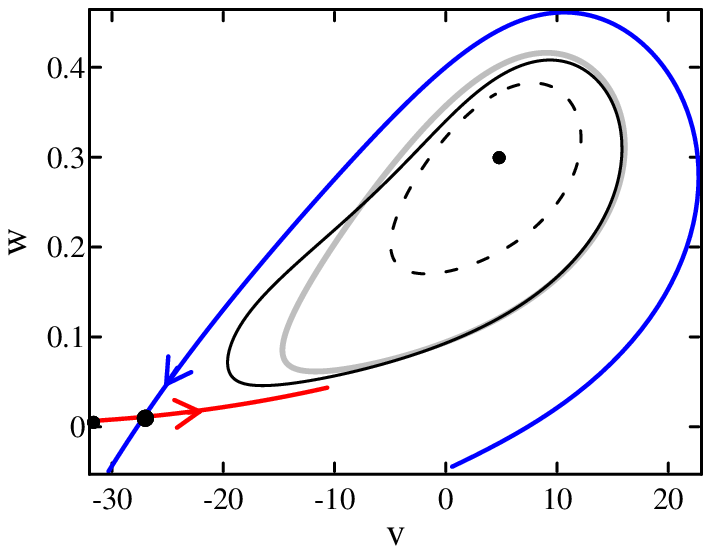} &
        \includegraphics*[bb=0.25in 0in 3in 2.25in,scale=0.65]{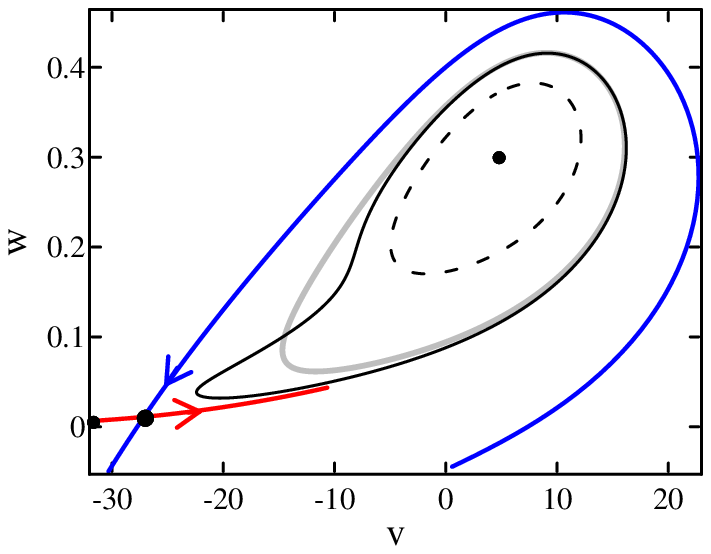} \\
        $t=15$ & $t=20$ & $t=25$ \\
        \includegraphics[bb=0in 0in 3in 2.25in,scale=0.65]{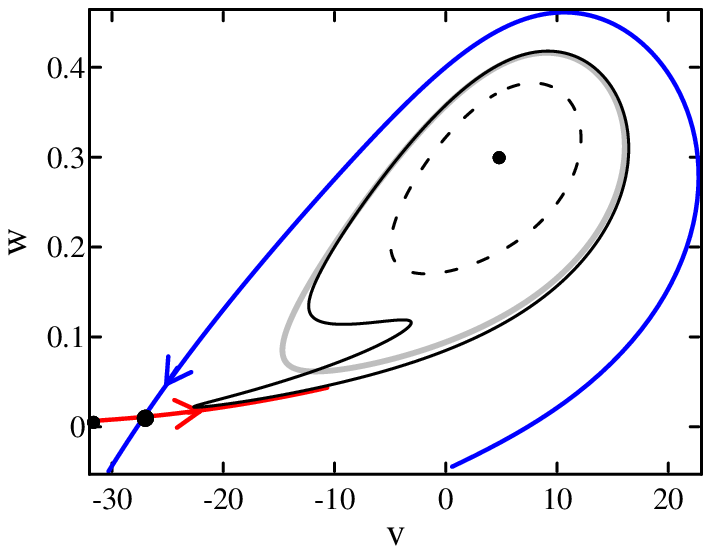} &
        \includegraphics*[bb=0.25in 0in 3in 2.25in,scale=0.65]{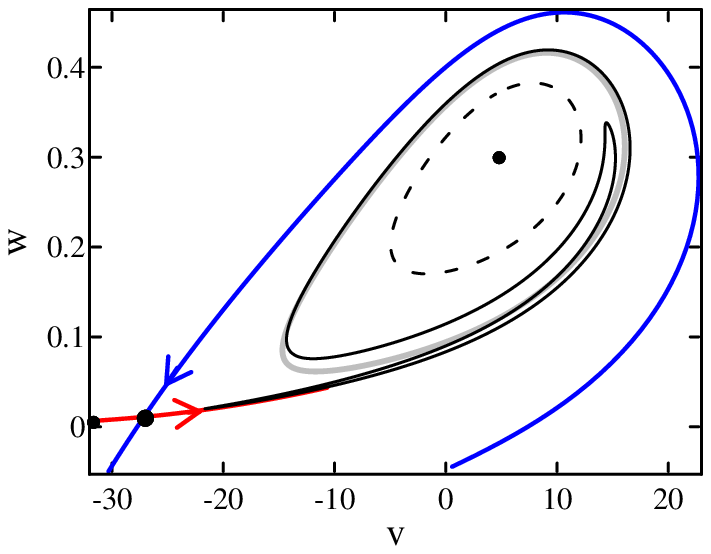} &
        \includegraphics*[bb=0.25in 0in 3in 2.25in,scale=0.65]{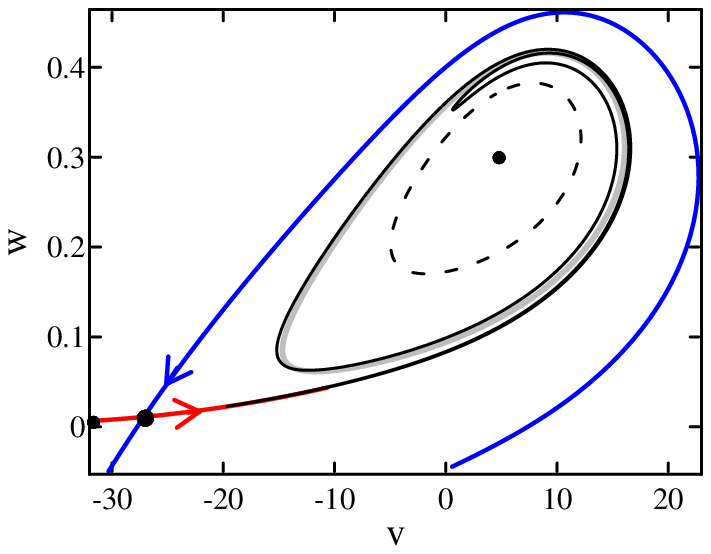} \\
      \end{tabular}
      
    \end{scriptsize}
  \end{center}
  
  \caption{Shear-induced folding in the Morris-Lecar model.  Here, the
    stable cycle is given one kick with $A=-2$, then allowed to relax
    back to the cycle.  The limit cycle $\gamma$ is shown in gray; also
    shown are the 3 fixed points (dots), the stable and unstable
    manifolds of the saddle, and the unstable cycle (dashed curve).}

  \label{fig:ml-fold}
\end{figure}

\medskip \noindent
{\em Simulations to detect shear-induced chaos}

\smallskip\noindent
As noted in Sect.~\ref{sect:going-forward}, by far the most direct way to detect the presence of
shear-induced chaos is to plot the images of $\kappa(\gamma)$ under
the unperturbed flow, and to see if folds develop in time. A sample ``movie" is
shown in Fig.~\ref{fig:ml-fold}. As one can see, by $t=10$, a ``tail" has developed: some
points in this tail are evidently stuck near the saddle, while some other points,
evidently $\Phi_t$-images of points kicked inside the unstable cycle, have 
remained inside, and at $t=10$ they are beginning to gain on points in the
tail in terms of their angular position (or phase) around $\gamma$.
At $t=15$, these points have overtaken those in the tail, and the difference
is further exaggerated in the last two frames. One would conclude that
for the parameters shown, the system very likely has shear-induced chaos.

Because this step can be done quickly for the ML model,
  we use it to locate parameters with the desired
properties. The kick size used in Fig.~\ref{fig:ml-fold} is $A=-2$,
which is quite reasonable physiologically: it takes at least 10-15 kicks
of this size to push a neuron over threshold.  Fig.~\ref{fig:ml-fold}
also tells us that it takes on the order of 15 units of time for the
fold to begin to form, so that for a periodically-kicked system to
produce chaos, the kick period should probably be upwards of 20 units of
time. (Kicks delivered too frequently may also drive points to the left
of the stable manifold of $p$; once that happens, it will end up near
the left sink.)
  
\begin{figure}
  \begin{center}
    \includegraphics[bb=0in 2.2in 5in 4.5in,scale=0.8]{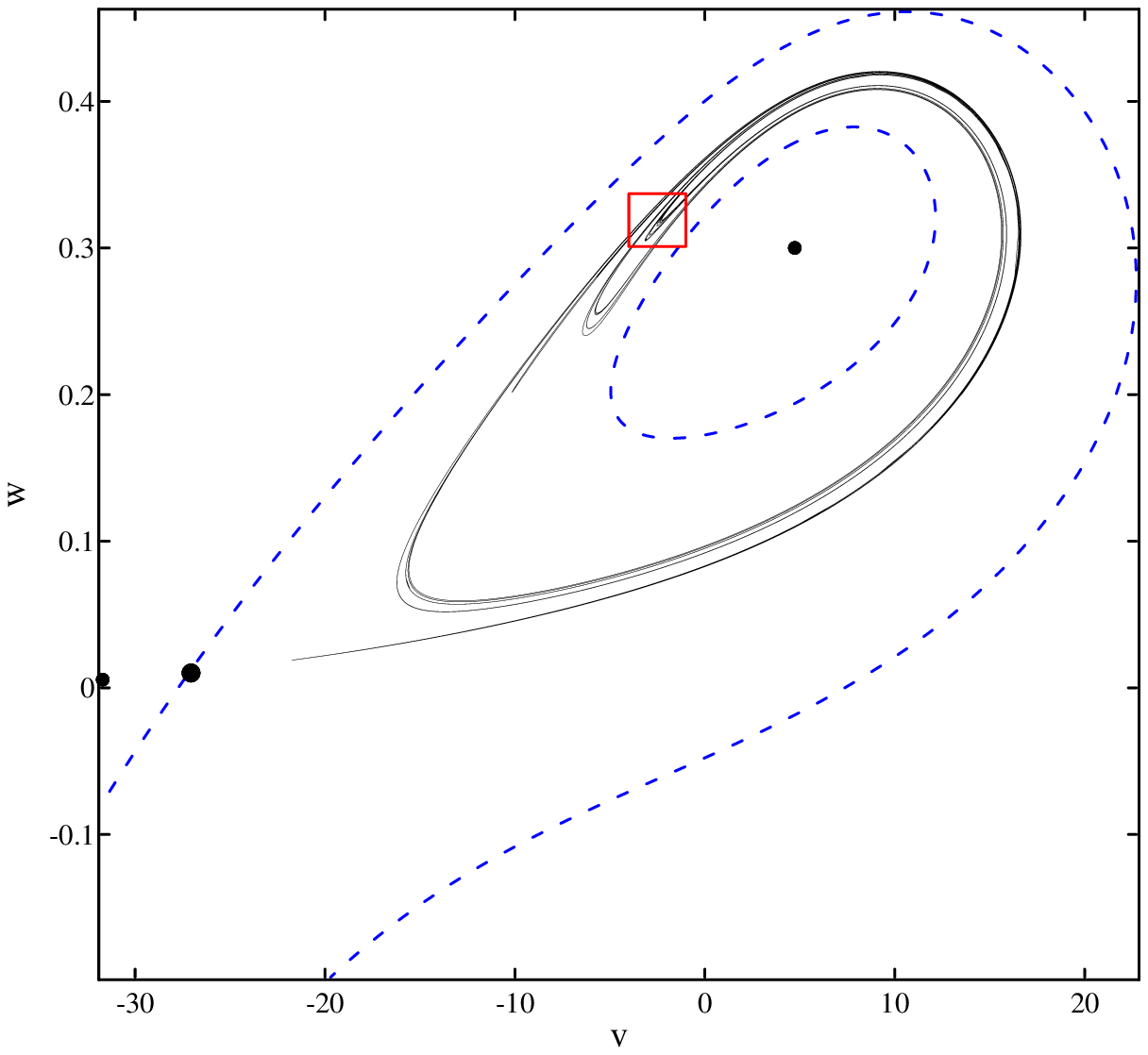}~~~~~~
    \includegraphics*[bb=1300 1722 1437 1845]{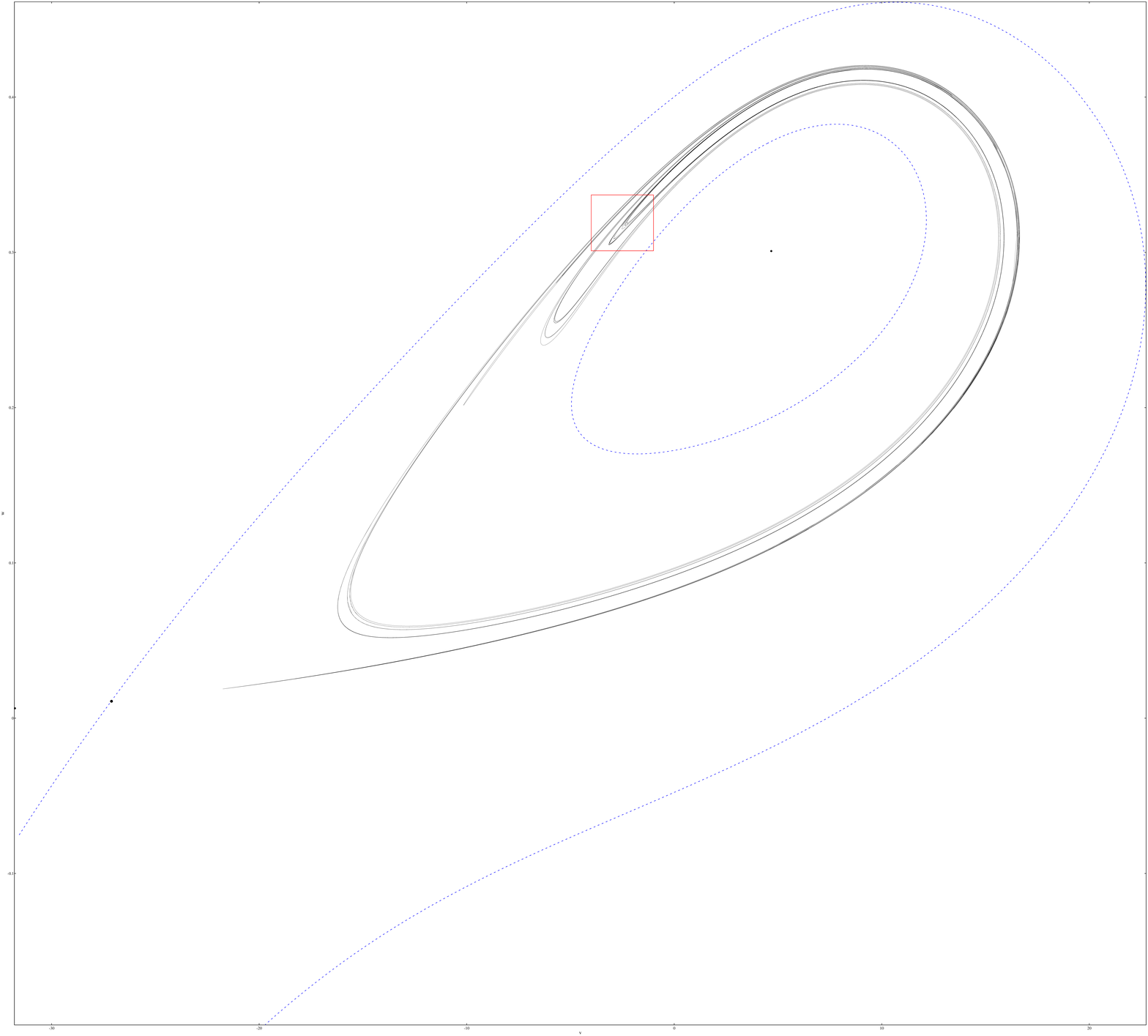}\\[1.7in]
  \end{center}

  \caption{A strange attractor created by shear-induced chaos.  The
    dashed curves show the basin boundary of the limit cycle $\gamma$~.
    Here $A=-2, T=27$.
  }
  \label{fig:attractor}
\end{figure}

Fig.~\ref{fig:attractor} shows the strange attractor that results from a periodically-kicked regime.

\medskip \noindent
{\em Lyapunov exponents}

\begin{figure}
  \begin{center}
    \includegraphics[bb=0in 0in 3in 2.5in,scale=0.66]{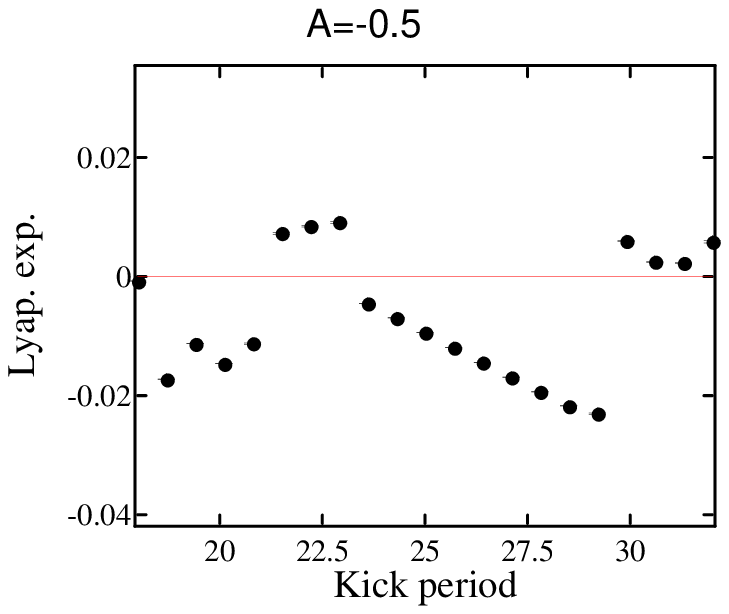}
    \includegraphics*[bb=0.25in 0in 3in 2.5in,scale=0.66]{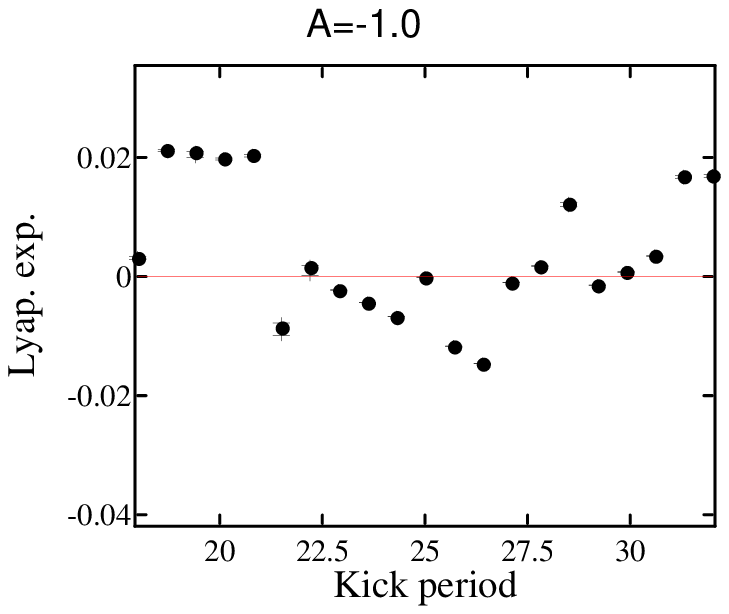}
    \includegraphics*[bb=0.25in 0in 3in 2.5in,scale=0.66]{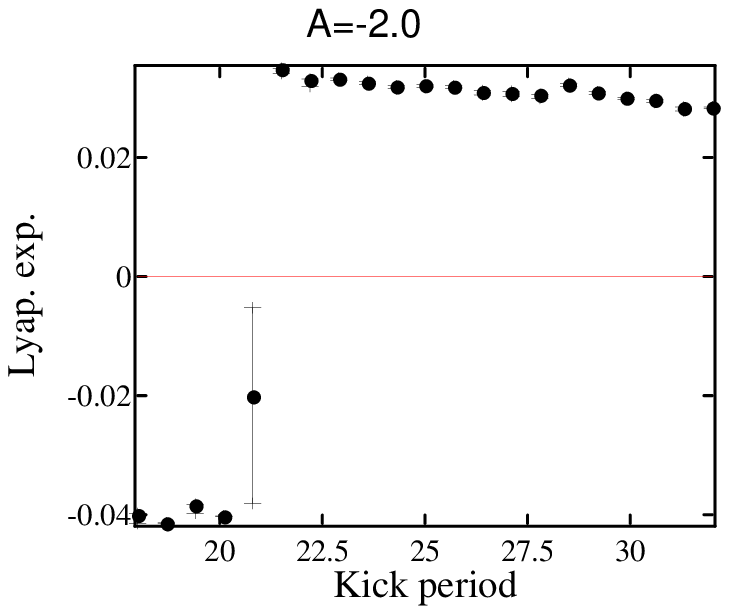}\\
    (a) Lyapunov exponents\\[3ex]
    \begin{tabular}{cc}
      Reliable ($\lmax\approx-0.02$)&Unreliable ($\lmax\approx0.035$)\\
      \includegraphics[bb=0in 0in 3.8in 1.25in,scale=0.75]{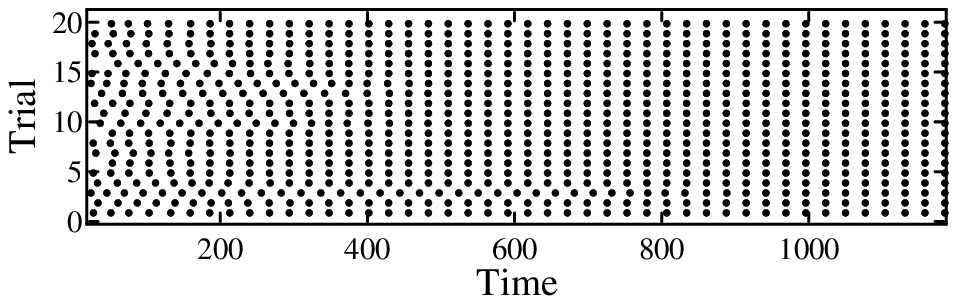}&
      \includegraphics*[bb=0.45in 0in 4in 1.25in,scale=0.75]{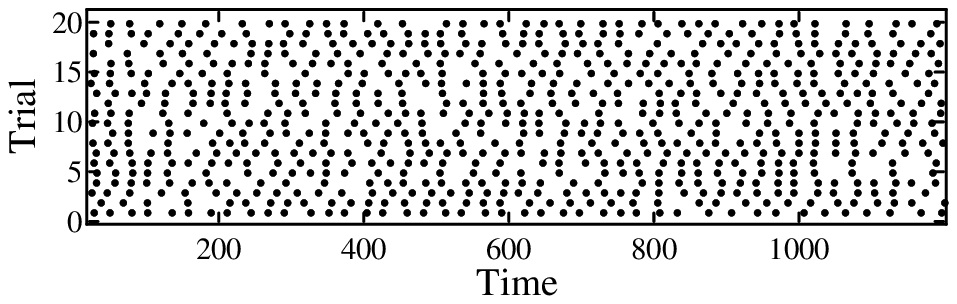}
    \end{tabular}\\[-2ex]
    (b) Spike times over repeated trials
  \end{center}
  \caption{Lyapunov exponents and reliability of the ML system.  In
      (a), we plot the Lyapunov exponents.  In (b), we plot spike times
      generated by reliable (left) and unreliable (right) systems over
      repeated trials, with random initial conditions sampled from a
      neighborhood of $\gamma$.  The exponents in (a) are computed as
    follows: For each choice of kick amplitude and kick period, 6 random
    initial conditions are used to estimate exponents via long-time
    simulations.  The max and min are treated as outliers and discarded;
    the remaining are used to form error bars.  The dot marks the
    median.}
  \label{fig:lyap-exps}
\end{figure}

\smallskip
\noindent
Let us now explore the chaotic behavior more systematically.
Fig.~\ref{fig:lyap-exps}(a) shows the largest Lyapunov exponent $\lmax$ as a
function of the kick period $T$, for a variety of kick amplitudes.  More
precisely, given a phase point $(v_0, w_0)$ and a tangent vector $\eta_0$, we let 
$\eta(t)$ be the solution of the variational equation for the kicked system,
and define
$$\lmax (v_0, w_0; \eta_0) = \lim_{t\to \infty} \frac{1}{t} \log |\eta(t)|\ .
$$
It is a standard fact that for each $(v_0, w_0)$, 
$\lmax (v_0, w_0; \eta_0)$ is equal to the largest Lyapunov
exponent except for $\eta_0$ in a lower dimensional subspace (see, e.g., \cite{eckrue}).
In many (but not all) of the cases considered, this number is independent
of the choice of $(v_0,w_0)$.

For small $A$, the exponents are predominantly negative, just as we
would expect: the dynamical landscape is characterized by sinks and
saddles.  As $A$ increases, the tendency to form positive exponents
becomes greater, so that for $A=-2$ and sufficiently large $T$, most
exponents sampled are positive, confirming the strong form of (sustained
and observable) chaos discussed in Sect.~\ref{wang-young}.  Note,
however, that this is only a tendency: the fluctuations seen in the
Lyapunov exponents as we vary $T$ are likely real, and reflect the
competition between the ``horseshoes and sinks" and ``strange
attractors" scenarios.

As explained in Sect.~\ref{sect:neurosci}, Lyapunov exponents are useful
as probes of neuronal reliability.  This is illustrated in
Fig.~\ref{fig:lyap-exps}(b).  The left panel there shows the response of
a system with $\lmax<0$: clearly, after some transients, the response of
the system is the same across all the trials.  In the right panel, we
have $\lmax>0$, and the resulting spike times show significant
variability across trials.  Moreover, this variability will persist in
time due to the presence of sustained chaos.

\medskip
We finish with the following two remarks: (1) While we have focused on the
periodically kicked case, similar-sized kicks that arrive at random times
that are, on average, sufficiently far apart will lead to {\it time-dependent
strange attractors} for much the same reasons as we have discussed. 
We will not pursue this here, but refer the reader to \cite{LY-nonlin,LSY12}.
(2) Notice that the saddle $p$,
which is the root cause of the chaos in the kicked system in the sense that
it is responsible for scrambling the order (or phases)
of the points on $\kappa(\gamma)$, 
does not lie that close to the limit cycle. Not only would phase reduction methods
miss this effect, but the PRC will offer no hint at all that any breakdown has
occurred.

\section{Firing rates and interspike intervals}
\label{sect:stuck}

We demonstrate in this section that, as asserted in item (2) of Sect.~\ref{sect:when-prc-works}, 
``sticky" sets in the boundary of ${\rm Basin}(\gamma)$ can lead to discrepancies
between the dynamical behavior of the full model and that deduced from the
infinitesimal PRC. 
Specifically, we will present numerical data
to show in the example considered that
{\em firing rates for the full model are significantly lower} and 
{\em interspike intervals (ISIs) are longer and more spread out}
than those predicted by the PRC. 

\medskip
For the unperturbed flow, we continue to use the example from the last
section, with the same parameters. Recall that this system is in the
``homoclinic regime" of ML, with a limit cycle we call $\gamma$.
Shown in dashed lines in Fig.~\ref{fig:attractor} are part of $\partial({\rm Basin}(\gamma))$, 
the boundary
of the basin of attraction of $\gamma$: one component is the stable manifold
$W^s$ of $p$; the other is a repelling periodic orbit. As discussed in
Sect.~\ref{sect:sic-in-ml}, all orbits close enough to 
$W^s$ will follow it to $p$ and remain there for some time before moving
away, $\{p\}$ is a candidate ``sticky set"  in this example. 
The repelling periodic orbit is in some sense also a sticky set, especially
if the expansion away from the orbit is weak (it is, since it has just emerged
from a Hopf bifurcation). Indeed orbits that come close to it may cross over
to the other side and get pushed toward the sink.

Instead of periodic forcing, here we drive the oscillator with white
noise in addition to the steady current $I_0$, i.e., we let $I(t) = I_0
+ \beta\cdot C_m\cdot\dot{W}~$,\footnote{The constant $C_m$ is the
  membrane capacitance.  With this scaling, $\beta$ has the dimension
  ${\rm voltage} / \sqrt{\rm time}$~; this makes its magnitude relative
  to the membrane voltage easier to assess.}  and consider the {\it
  random dynamical system} that results from looking at one realization
of $\dot{W}$ at a time.  For the phase reduced model, it is natural to regard
the system as producing one spike as it passes through a marked point on
the cycle.  For the full ML model, it is necessary to fix an artificial
definition of what it means for the system to ``spike": we define a
reset to be when the voltage falls below -10, and say the neuron
``spikes" when (after a reset) the voltage rises above +12.5. See
Fig.~\ref{fig:attractor} for how this corresponds to the location of the
limit cycle.

\begin{figure}
  \begin{center}
    \begin{tabular}{cc}
      \includegraphics[bb=0in 0in 3.25in 2.5in,scale=0.6]{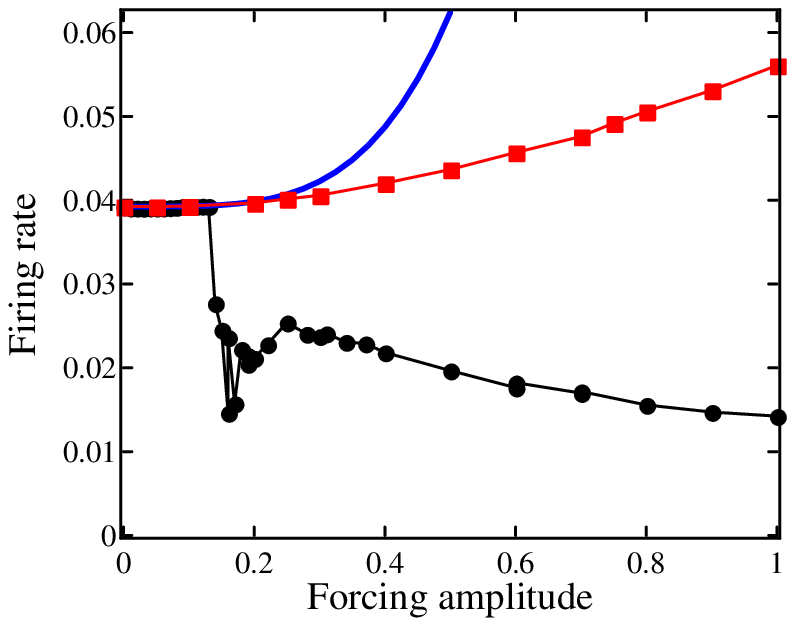} &
      \includegraphics[bb=0in 0in 3in 2.75in,scale=0.6]{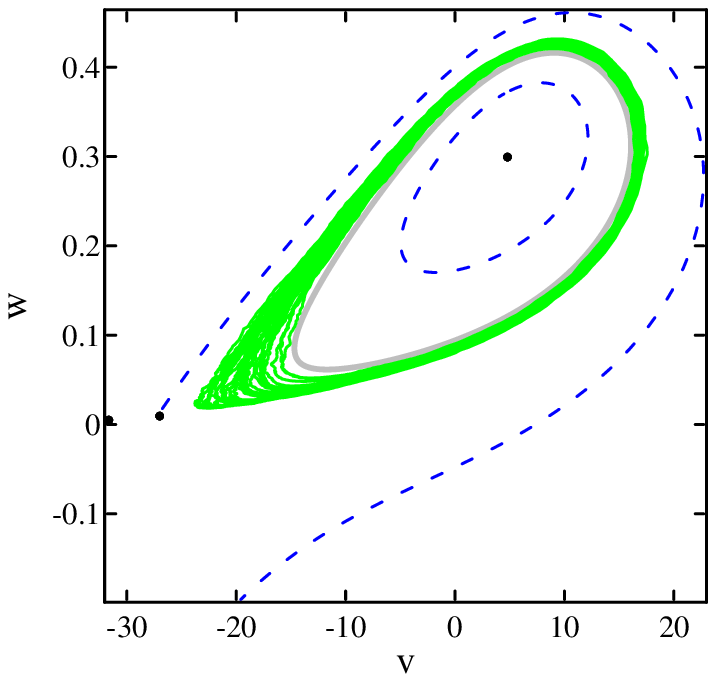}\\
      (a) & (b) \\
    \end{tabular}

  \end{center}
  
  \caption{Firing rate statistics of the stochastically-forced ML model.
    In (a), we compare of firing rates for the full ML system and its
    phase reduction driven by white noise.  Black dots: Full ML.  Red
    squares: Empirical phase reduction firing rate.  Solid blue curve:
    Perturbative formula for PRC firing rate (see
    Eq.~(\ref{eq:prc-firing-rate})).  Panel (b) shows a long sample path
    for $\beta=0.2$~.}
  \label{fig:hom-firing-rates}
\end{figure}

\heading{Comparison of firing rates.}   The
results are shown in Fig.~\ref{fig:hom-firing-rates}(a).  Plotted are
firing rates as a function of the drive amplitude $\beta$ for three
different systems: (i) the full ML system; (ii) the firing rate of the
phase reduction, computed empirically; and (iii) the firing rate of the
phase reduction as computed by the perturbative
formula~(\ref{eq:prc-firing-rate}) of Ly and Ermentrout.  We plot both
(ii) and (iii) because Eq.~(\ref{eq:prc-firing-rate}) is itself an
approximate result based on the phase reduction (\ref{eq:prc-sde}),
valid only for small $\beta$. 
As one would expect, for smaller forcing amplitudes ($< 0.1$), all three
agree.  As $\beta$ increases, the perturbative formula tracks the
empirical firing rate of the phase reduction fairly well, but neither of
these PRC-based predictions capture the dramatic drop in firing rate of
the full system occurring around $\beta=0.2$.  Simulations 
(an example of which is shown in Fig.~\ref{fig:hom-firing-rates}(b))
show that the precipitous drop in firing rate in this case has to do
with trajectories spending more time near the saddle $p$.

\begin{figure}

  \begin{center}
    \includegraphics*[bb=0.25in 0in 3in 2.5in,scale=0.7]{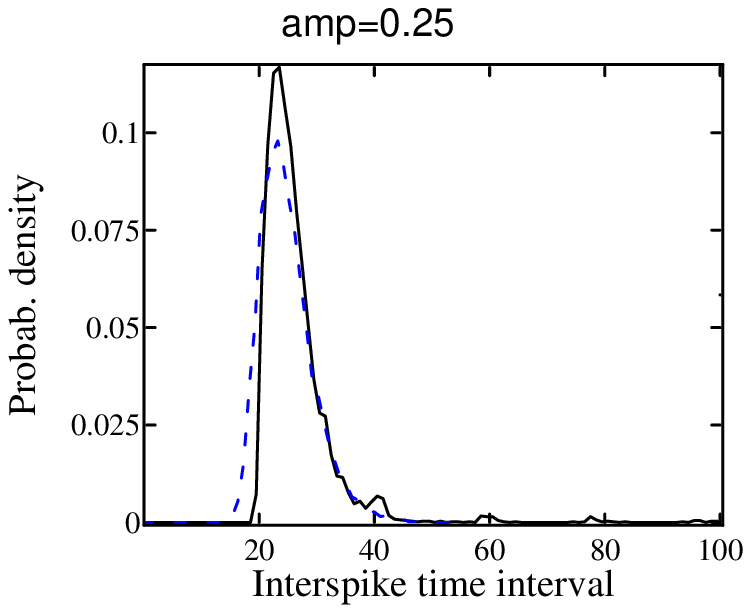}
    \includegraphics[bb=0in 0in 3in 2.5in,scale=0.7]{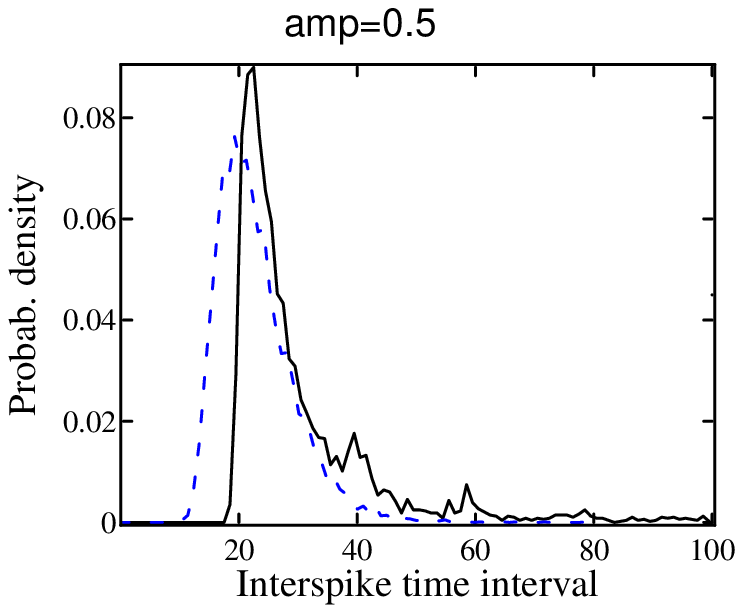}%
    \includegraphics*[bb=0.25in 0in 3in 2.5in,scale=0.7]{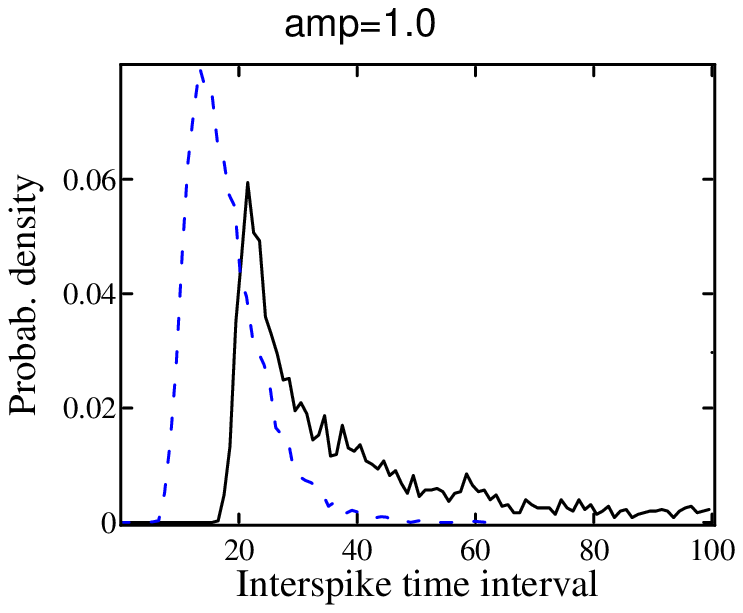}
  \end{center}
  
  \caption{interspike interval (ISI) distributions for various
    $\beta$.  Solid black lines: ISI histograms for the full ML system.
    Dashed blue lines: ISI histograms for the phase reduction
    (\ref{eq:prc-sde}).}
  \label{fig:hom-isi}
\end{figure}

\heading{Comparison of ISI distributions.}  Fig.~\ref{fig:hom-isi} shows the
numerically computed ISI distributions for the full ML model and for the
phase reduction (\ref{eq:prc-sde}).\footnote{We note that \cite{LE} also
  contains an explicit small-$\beta$ expansion of the ISI distribution.
  We do not use it in Fig.~\ref{fig:hom-isi} because it is not
  applicable except (possibly) for $\beta=0.05$~.}  For small noise, we
see the full ML system and phase reduction agree fairly well: the ISI
distribution is concentrated around the period of the cycle (about
25.2), and the shape is roughly gaussian. As noise amplitude increases,
we see in the full ML system (i) a broadening of the ISI distribution,
and (ii) an overall shift toward larger ISIs.  Neither of these effects
are captured by the PRC, as they involve structures a finite distance
away from $\gamma$~.

Finally, to compare the tails of the ISI distributions, here are the
fraction of ISIs greater than $2 \times\mbox{(cycle period)}$:
\begin{center}
  \begin{tabular}{c|c|c}
    $\beta$ & Full ML & PRC \\\hline
    0.25 & 0.0225 & 0.0003 \\
    0.5 & 0.127 & 0.0032 \\
    1.0 & 0.322 & 0.0013 \\
  \end{tabular} \\
\end{center}
Notice that the PRC systematically {\em under-predicts} the probability
of a long interspike interval, consistent with the ``trapping'' effect
of nearby invariant structures.

\section{Altered spiking patterns in bistable systems}
\label{sect:bistable}

In this last section, we present a simple example in which the
perturbation takes an orbit outside of ${\rm Basin}(\gamma)$, the
scenario described in item (1) of Sect.~\ref{sect:when-prc-works}.  It
is no surprise that PRC predictions would break down here.  Perhaps the
lesson to take away from this example is that even a weak drive can lead
to spike patterns that are nontrivially different.

\begin{figure}

  \begin{center}
    \includegraphics[bb=0in 0in 4in 3.75in,scale=0.6]{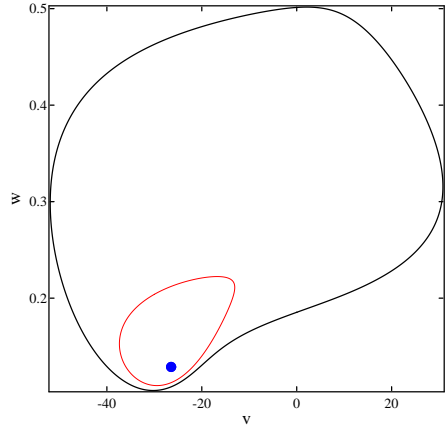}
  \end{center}
  
  \caption{ML phase portrait in the Hopf regime.  Shown are a limit
    cycle surrounding an unstable cycle, which in turn surrounds a sink.
    Here, $I_0=90$, corresponding to an unforced period of about 102
    ms.}
  \label{fig:hopf-portrait}
\end{figure}

\medskip
For the unperturbed system, we consider here parameters of ML that put
it in a ``Hopf regime", with $I_0 \approx 90$.  Precise parameters used
are given in the Appendix, and some relevant dynamical structures are
shown in Fig.~\ref{fig:hopf-portrait}.  In this regime, the system has a
limit cycle, which we call $\gamma$.  Each time an orbit passes near the
right-most point of $\gamma$, we think of it as producing a ``spike".
Notice that this system is {\it bistable}: inside $\gamma$ there is a
sink, the basin of attraction of which is separated from the basin of
$\gamma$ by a repelling periodic orbit (from a Hopf bifurcation). This
repelling periodic orbit is very close to $\gamma$, but since it is a
positive distance away, it will not show up in PRC considerations.

\begin{figure}
  \begin{center}
    \begin{small}
      Full ML, $\beta = 0.2$ \\
      \includegraphics*[bb=0in 0.22in 6.25in 1.25in,scale=0.95]{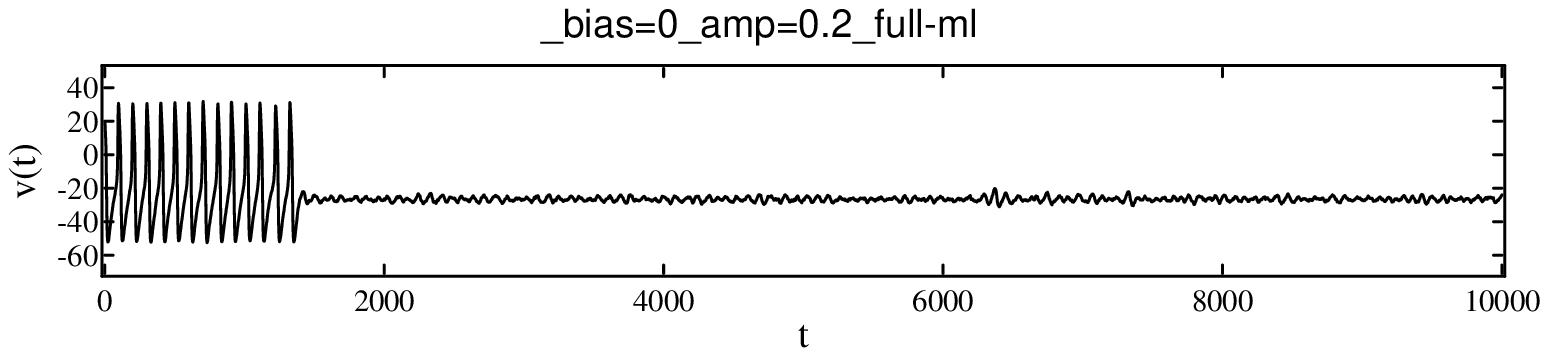}\\[2ex]
      Full ML, $\beta = 0.4$ \\
      \includegraphics*[bb=0in 0.22in 6.25in 1.25in,scale=0.95]{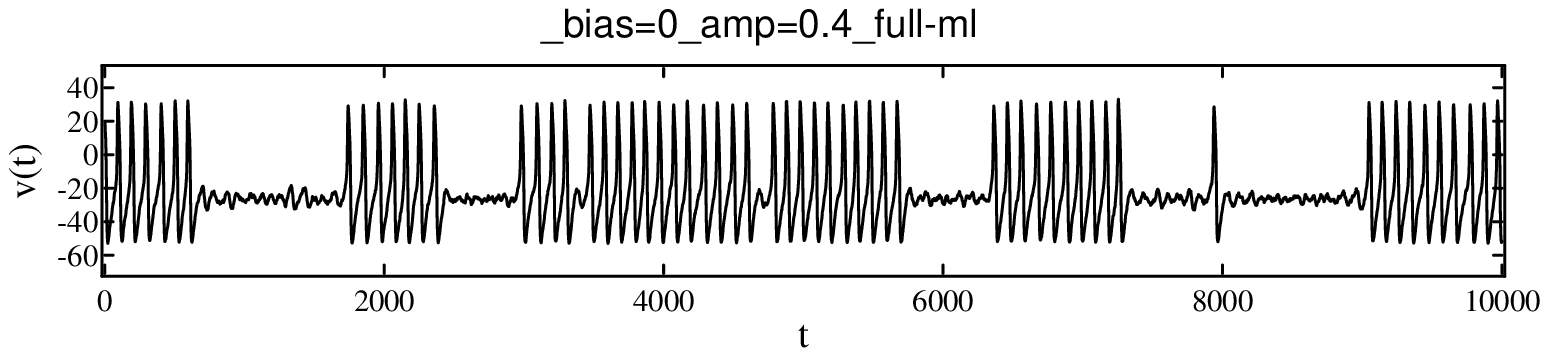}\\[2ex]
      PRC, $\beta = 0.4$ \\
      \includegraphics*[bb=0in 0.22in 6.25in 1.25in,scale=0.95]{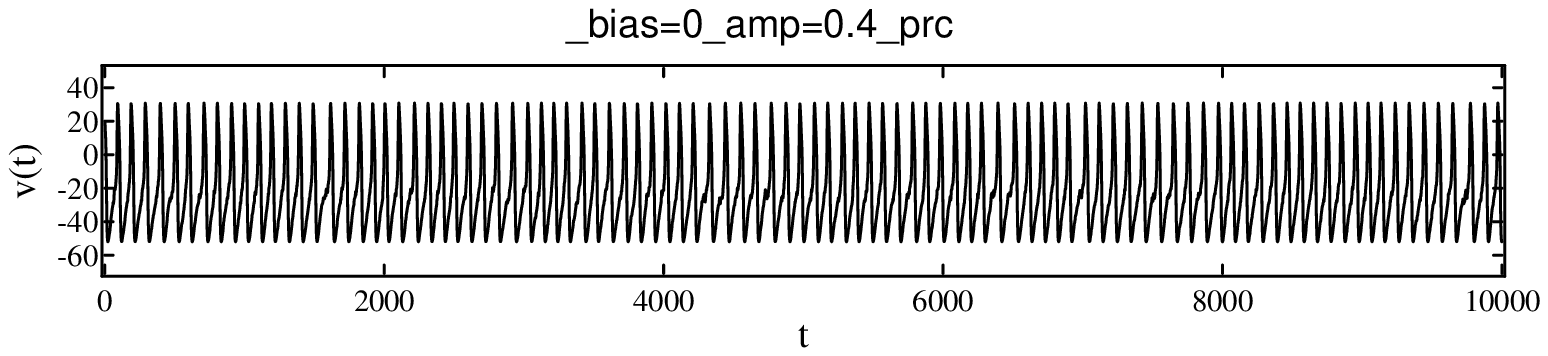}\\[-2ex]
      $t$\\

    \end{small}
  \end{center}

  \caption{Voltage traces for the ML system in Hopf regime.  The top two
    panels show simulation using the full ML model with the indicated
    level of stochastic forcing.  The bottom panel shows the PRC using
    $\beta=0.4$.
    Here, $I_0=90$.}
  \label{fig:hopf-traces}
\end{figure}

We now consider perturbations to the ML model above.  As in
Sect.~\ref{sect:stuck}, we again use a white-noise perturbation, i.e.,
$I(t) = I_0 + \beta\cdot C_m\cdot \dot{W}~.$
Fig.~\ref{fig:hopf-traces} shows the resulting voltage traces of the
full model and PRC predictions (computed at $I_0=90$).  The top panel
shows the full ML simulation with $\beta=0.2$ (a relatively weak
forcing), starting the trajectory on $\gamma$~.  Since the basin of the
sink, i.e., the unstable periodic orbit shown in
Fig.~\ref{fig:hopf-portrait}, is so close to the limit cycle $\gamma$, a
trajectory following $\gamma$ can easily get pushed into the basin, and
then attracted to the sink.  This can happen even with very small
$\beta$.  The sink itself, however, is a little farther from the
boundary of its basin, so it is more difficult for a trajectory near the
sink to escape under weak $\beta$.  This is why for weak noise like
$\beta=0.2$, it is easy to observe a transition from $\gamma$ to the
sink, but not easy to see transition in the reverse direction.

For $\beta$ a little larger, e.g., $\beta=0.4$, the trajectory jumps
back and forth more readily: it alternately follows (roughly) the limit
cycle $\gamma$ and stays near the sink, switching between the ``spiking"
and ``quiescent" modes at somewhat random times.  Not surprisingly, in
the phase-reduced model (bottom panel), these perturbations do not have
a significant impact, and the spiking ranges from completely regular to
slightly irregular in the second case due to the effect of the noise.
PRC voltage traces for $\beta=0.2$ (not shown) are quite similar.

Finally, we note that in this bistable situation, a neuron can exhibit
substantial sub-threshold activity.  The PRC underestimates the extent of
such activity.

\section*{Conclusions}

This paper compares two ways of evaluating an oscillator's response to
perturbations: phase reductions versus analysis of the full model. We have
found that the infinitesimal PRC, 
which has the virtue of being both
 straightforward to execute and reducing model dimensionality, produces
regular oscillatory behavior even when the full model does not.
Specifically, we presented examples from the Morris-Lecar neuron model to show
that
\begin{enumerate}

\item Periodic kicking of the ML system can lead to unreliable response
in the full model
 via the mechanism of shear-induced chaos, contrary to PRC predictions.

\item When stochastically driven, stickiness
of nearby invariant structures can lead to lower firing rates
and longer ISIs compared to PRC predictions.

\item The forcing need not be strong to bring about serious discrepancies
in firing patterns between full and phase-reduced models.

\end{enumerate}
Moreover, in all the situations examined, the phase reduction itself offers no hint
that any breakdown has occurred.

In terms of neuronal response, our results have the following interpretation:
Under certain conditions, PRCs may overestimate
spike-time reliability and firing rates; they may also underestimate the
mean and variance in interspike intervals, and have a tendency to
downplay sub-threshold activity.  Caution needs to be exercised when
interpreting results that come from phase reduction arguments, especially
for systems near bifurcation points. 

While we have focused on the ML model for illustration, the
geometric ideas we used are quite general, and we expect similar
phenomena to occur in a variety of settings that involve rhythms and oscillations.

Finally, given that pure phase reductions cannot always capture the
behavior of high dimensional nonlinear oscillators, are there
alternative methods that give better characterizations?  One classical
technique, which has seen relatively little application in biological
modeling, is that of moving-frame coordinates for the analysis of
periodic orbits.  This is nicely described in Hale \cite{Hale69} and
recently advocated in a neuroscience context by Medvedev
\cite{Medvedev11}.  In essence this offers a coordinate transformation
to a phase-\textit{amplitude} system that allows one to track the
evolution of distance from the cycle as well as phase on the cycle.  We
are currently developing this approach and using it to better understand
the three points listed above.  This, and a framework for understanding
the dynamics of weakly coupled phase-amplitude models, will be presented
elsewhere.

\paragraph{Acknowledgements.}
KKL is supported in part by the US National Science Foundation (NSF)
through grant DMS-0907927.  KCAW and SC acknowledge support from the
CMMB/MBI partnership for multiscale mathematical modelling in systems
biology - United States Partnering Award; BB/G530484/1 Biotechnology and
Biological Sciences Research Council (BBSRC).  LSY is supported in part
by NSF grant DMS-1101594.

\section*{Appendix.  Morris-Lecar model details}

Here we briefly summarize the details of the ML model used in this
paper.  The interested reader is referred to \cite{ER,ET} for more
details.

Recall the ML equations
\begin{displaymath}
  \begin{array}{rcl}
    C_m~\dot{v} &=& I(t) - g_{\rm leak}\cdot(v-v_{\rm leak}) -
    g_K~w\cdot(v-v_K) - g_{\rm Ca}~m_\infty(v)\cdot(v-v_{\rm Ca}) \\[2ex]
    \dot{w} &=& \phi\cdot\big(w_\infty(v) - w\big)/\tau_w(v)~.
  \end{array}
\end{displaymath}
As explained in Sect.~\ref{sect:neurosci}, the ML model tracks the
membrane voltage $v$ and a gating variable $w$.  The constant $C_m$ is
the membrane capacitance, $\phi$ a timescale paramter, and $I(t)$ an
injected current.

Spike generation depends crucially on the presence of voltage-sensitive
ion channels that are permeable only to specific types of ions.
The ML equations include just two ionic currents, here denoted calcium
and potassium.  The voltage response of ion channels is governed by the
$w$-equation and the auxiliary functions $w_\infty$~, $\tau_w$~, and
$m_\infty$~, which have the form
\begin{displaymath}
 \begin{array}{rcl}
   m_\infty(v) &=& \frac12\big[1+\tanh\big((v-v_1)/v_2\big)\big]~, \\[1.5ex]
   \tau_w(v) &=& 1/\cosh\big((v-v_3)/(2v_4)\big)~,\\[1.5ex]
   w_\infty(v) &=& \frac12\big[1+\tanh\big((v-v_3)/v_4\big)\big]~.
 \end{array}
\end{displaymath}
The function $m_\infty(v)$ models the action of the relatively fast
calcium ion channels; $v_{\rm Ca}$ is the ``reversal'' (bias) potential
for the calcium current and $g_{\rm Ca}$ the corresponding conductance.
The gating variable $w$ and the functions $\tau_w(v)$ and $w_\infty(v)$
model the dynamics of slower-acting potassium channels, with its own
reversal potential $v_{\rm K}$ and conductance $g_{\rm K}$~.  The
constants $v_{\rm leak}$ and $g_{\rm leak}$ characterize the ``leakage''
current that is present even when the neuron is in a ``quiescent''
state.  The forms of $m_\infty$~, $\tau_w$~, and $w_\infty$ (as well as
the values of the $v_i$) can be obtained by fitting data, or reduction
from more biophysically-faithful models of Hodgkin-Huxley type
(see, e.g., \cite{ET}).

The precise parameter values used in this paper are summarized in
Table~\ref{tab:params}.  These are obtained from \cite{ET}.

\begin{table}
  \begin{center}
    \begin{tabular}{c|c|c}
      Parameter & Homoclinic regime & Hopf regime \\\hline
      $I_0$ & 39.5 & 90 \\
      $C_m$ & 20.0 & 20.0 \\
      $g_{\rm Ca}$ & 4.0 & 4.4 \\
      $g_K$ & 8.0 & 8.0 \\
      $g_{\rm leak}$ & 2.0 & 2.0 \\
      $v_K$ & -84.0 & -84.0 \\
      $v_{\rm leak}$ & -60.0 & -60.0 \\
      $v_{\rm Ca}$ & 120.0 & 120.0 \\
      $\phi$ & 0.23 & 0.04 \\
      $v_1$ & -1.2 & -1.2 \\
      $v_2$ & 18.0 & 18.0 \\
      $v_3$ & 12.0 & 2.0 \\
      $v_4$ & 17.4 & 30.0 \\
    \end{tabular}
  \end{center}
  \caption{ML parameter values used in this paper.}
  \label{tab:params}
\end{table}

\end{document}